\documentclass[rmp,nofootinbib,twocolumn]{revtex4}
\usepackage{graphicx}
\ifnum\lefthyphenmin<2\lefthyphenmin=2\fi
\ifnum\righthyphenmin<2\righthyphenmin=2\fi
\newcommand{\cQ}[0]{{\cal Q}}
\bibpunct[, ]{(}{)}{;}{a}{,}{,}
\begin{document}
\title{\bf Force-induced denaturation of RNA}
\author{Ulrich Gerland}
\author{Ralf Bundschuh}
\author{Terence Hwa}
\affiliation{Department of Physics, University of California at San Diego\\
La Jolla, California 92093-0319}
%
%
%
\begin{abstract}
We describe quantitatively a RNA molecule under the influence of an
external force exerted at its two ends as in a typical single-molecule
experiment. 
Our calculation incorporates the interactions between nucleotides by using 
the experimentally-determined free energy rules for RNA secondary structure 
and models the polymeric properties of the exterior single-stranded regions
explicitly as elastic freely-jointed chains. 
We find that in spite of complicated secondary structures,
force-extension curves are typically smooth in quasi-equilibrium. 
We identify and characterize two sequence/structure-dependent mechanisms that, 
in addition to the sequence-independent entropic elasticity of the exterior 
single-stranded regions, are responsible for the smoothness. 
These involve compensation between different structural elements on 
which the external force acts simultaneously, and contribution of 
suboptimal structures, respectively. 
We estimate how many features a force-extension curve recorded in 
non-equilibrium, where the pulling proceeds faster than rearrangements in 
the secondary structure of the molecule, could show in principle. 
Our software is available to the public through a `RNA-pulling server'. 
\end{abstract}
\maketitle
\section{Introduction}
In recent years, single-molecule experiments employing optical tweezers,  
atomic force microscopy, and other techniques have successfully probed 
basic physical properties of biomolecules through the application of 
forces in the pN range (see, e.g., 
\citet{smith,mehta,bockelmann,heslot,rief99,rief97,yang}). 
Both, simple elastic properties of the polymers (such as persistence length 
and longitudinal elasticity) and structural transitions (e.g. unfolding 
of protein domains) were characterized by recording and analyzing 
force-extension curves (FEC's). 
For nucleic acids, a prominent experiment of the latter type is the 
`unzipping' of double-stranded DNA \citep{heslot,bockelmann}. 
The resulting FEC's display clear sequence-specific features (e.g. local 
maxima), which may be attributed to small regions of the sequence that 
are more strongly bound than their neighbors~\citep{heslot,siggia,lubensky}.
In contrast, long single-stranded DNA, which, like RNA, may fold into 
complicated branched structures by forming intra-strand basepairs, 
showed extremely smooth FEC's in a very recent experiment by 
\citet{bensimon}. 
Thus, depending on its structure, DNA may show a broad range of FEC's 
from very rugged to completely featureless. 
However, it is unclear {\em how} quantitatively the structure determines 
the outcome of the FEC measurement. 

Here, we address this question theoretically, focusing on the case 
of RNA and restricting ourselves to secondary structure 
(i.e. basepairing patterns only instead of full, tertiary structure).  
In this context, RNA seems to be a more interesting object than DNA, 
since RNA naturally occurs in many different and functionally important 
structures, while DNA is primarily found as a double strand. 
One may hope that pulling experiments generate
new insights into the RNA folding problem~\citep{tinoco}, including
the folding pathways~\citep{thirumalai,chen,isambert}. 
Also, force-induced denaturation of RNA is currently studied 
experimentally (C.~Bustamante and I.~Tinoco, private communiation).
The limitation to secondary structure allows us to draw upon the 
experimentally determined `free energy rules' for RNA secondary 
structure~\citep{freier,walter,mathews}, which yield minimum free 
energy structures that agree reasonably well with 
experimentally and phylogenetically determined ones~\citep{mathews}.
Furthermore, it permits us to employ and extend the efficient dynamic
programming algorithms~\citep{zuker,mccaskill,vienna} which can compute 
the exact partition function (including all possible secondary structures) 
and reconstruct the minimal free energy structures in polynomial time. 
Experimentally, the secondary structures may be probed in specific ionic 
conditions (e.g., those with only monovalent ions) such that the tertiary 
contacts are strongly disfavored (due to electrostatic repulsion of the 
sugar-phosphate backbone)~\citep{tinoco}.

\begin{figure}[tb]
  \begin{center}
  \includegraphics[width=7cm]{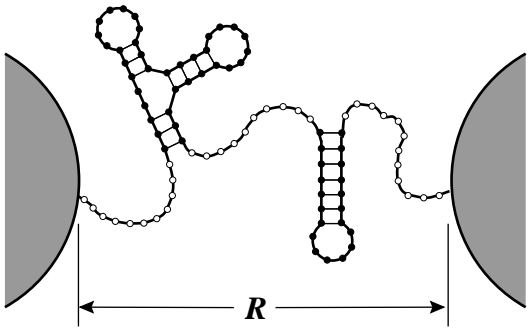}
  \end{center}
  \caption{
  Sketch of the pulling experiment considered in the text: the 
  two ends of a RNA molecule are attached to beads (shaded gray) 
  and held fixed at distance $R$, while the force $f$ acting on the 
  beads is measured.
  The open circles represent the open bases of the exterior single strands, 
  modeled here as elastic freely jointed chains.}
  \label{figsketch}
\end{figure} 

The type of experiment that we consider is sketched in Fig.~\ref{figsketch}. 
The distance $R$ between the two ends of an RNA molecule is held fixed, 
e.g., by attaching them to two beads whose positions are controlled by optical 
tweezers, and the force $f$ acting on the beads is recorded as a function 
of $R$. As long as the external change in force/extension is applied at a 
much slower time scale than that of structural transitions of the molecule,
the equilibrium FEC is measured. 
In the main part of the present article, we assume that this is always the 
case. Experimentally, this condition is usually checked by retracing the 
FEC (e.g., a hysteresis effect is a clear sign of a non-equilibrium 
situation). 

Besides the above-mentioned free energy parameters for RNA secondary 
structure, we need a polymer model for single-stranded RNA as 
input in order to make quantitative predictions of FEC's. 
To that end, we employ an elastic freely jointed chain model, which has 
been used to fit experimental FEC's of single-stranded 
DNA \citep{smith,mezard}. 
This introduces two polymer parameters, the Kuhn length characterizing the 
lateral rigidity, and the longitudinal elasticity, which is determined by 
the forces needed to stretch the chemical structure of the backbone. 
We estimate both from the experiments on DNA, so that we are 
left with no free parameters. 

We find that for different secondary structures with all other 
parameters (temperature, sequence length, etc.) fixed, the FEC's of RNA 
vary over a broad range from very rugged to very smooth. 
Apart from the entropic elasticity of the exterior single strand, which 
smoothens the features in the FEC independent of the secondary structure 
as already discussed by \citet{siggia}, there are two additional smoothing 
mechanisms. 
The first is a `compensation effect': the increase in the length of the 
exterior single strand upon opening of a structural element and the 
associated drop in the tension may be absorbed by rebinding of bases from 
the exterior single strand in other structural elements. 
The second is due to thermal fluctuations in the secondary structure, i.e. 
the contribution of suboptimal structures. 
We discuss both mechanisms and analyze the fluctuations in the FEC 
quantitatively. 
The equilibrium FEC's of typical (natural or random) RNA sequences are 
smooth and display no distinguishable signatures of individual structural 
elements opening. 
This is consistent with the experimental result of 
\citet{bensimon} for single-stranded DNA, but applies even for 
sequences with only a few hundred nucleotides, i.e. for much shorter 
sequences than used in their experiment. 

For the purpose of obtaining information on the structure of RNA, the 
measurement of equilibrium FEC's is therefore not very useful. 
More promising options include the measurement of the fluctuations about 
the equilibrium and non-equilibrium FEC's, where the pulling proceeds 
faster than (some of) the rearrangements in the structure. 
While the present approach is extended readily to include equilibrium 
fluctuations (Gerland, U., R. Bundschuh, and T. Hwa, in preparation), 
a quantitative treatment of the dynamics of force-induced 
denaturation of RNA presents a challenge to theoreticians. 

The organization of the paper is as follows. In the next section, we 
explain the details of our model and the way we calculate the FEC's. 
Readers interested in the results only should directly proceed to 
section III. The discussion in section IV explores the possibility 
of using experimental FEC's of appropriately designed sequences as an 
alternative way to determine the RNA free energy parameters. 
In addition, we estimate to what extent features may be expected 
in non-equilibrium FEC's. 
\section{Model and Methods}
We assume that the force $f(R)$ acting on the beads (see 
Fig.~\ref{figsketch}) is measured as a function of the fixed distance 
$R=|{\bf R}|$, where ${\bf R}$ denotes the end-to-end vector of 
the RNA molecule, and that $R$ is varied very slowly so that
thermal equilibrium is always maintained.
In practice, the force measurement requires a device acting as a
spring, hence the distance cannot be kept exactly constant.
However, we consider the situation where the stiffness of this
spring is much higher than that of the single-stranded RNA, which has 
already been pulled out. This condition could only be violated in the 
very early part of the pulling experiment, which is not the focus of 
the present investigation. 
We may therefore neglect the presence of the spring altogether, which 
amounts to working in the `fixed-distance ensemble'\footnote{In the 
`fixed-distance ensemble', only the average force is well-defined, 
whereas the fluctuations about the average diverge. 
This reflects the fact that it takes increasingly higher forces to 
compensate thermal fluctuations on shorter and shorter timescales, 
in order to keep the extension {\em exactly} fixed. 
Therefore, if one is interested in the fluctuations (of either the 
force or the extension), the external spring should not be neglected, 
which would amount to working in a mixed ensemble between 
`fixed-distance' and `fixed-force'.}. 
Another difference between our model and actual experiments is that we 
neglect the presence of additional spacer sequences, which are used to 
connect the RNA molecule to the force-measuring device (e.g. the beads). 
Again, we assume that they are stiffer than the liberated 
single-stranded RNA, since we are interested in the size of the 
features in the FEC, which are observable in an ideal measurement. 

The partition function at fixed extension, $Z_N(R)$, for a given
RNA sequence consisting of $N$ nucleotides, may be written as a sum
over the number $m$ of exterior open bases (as represented by open
circles in Fig.~\ref{figsketch}). 
For each $m$, the secondary structure contributes a factor
$\cQ_N(m)$ to the partition function, according to the free energy rules for
RNA/DNA secondary structure to be detailed shortly below.
This contribution needs to be weighted by the probability
$W({\bf R};m)$ that the chain of $m$ exterior open bases has
end-to-end vector ${\bf R}$, given by an appropriate polymer model 
for the single strand. Together, they yield
\begin{equation}
  \label{convolution}
  Z_N(R)=\sum_m \,\cQ_N(m)\,W({\bf R};m)\;.
\end{equation}
The normalization $\int\!{\rm d^3}\!R\,\,W({\bf R};m)\!\!=\!\!1$
assures that the integral of $Z_N(R)$ over space yields
the usual partition function $Z_N$ for $N$ nucleotides without
any external constraints.
Eq.~(\ref{convolution}) clearly separates the contribution of the
secondary structure, which is entirely contained in
$\cQ_N(m)$, from the contribution of 
the exterior single strand contained in $W({\bf R};m)$.
Note that the polymer properties of the {\em interior} single strands
(i.e. the single strands not subject to the external force)
are  contained in $\cQ_N(m)$ through the loop-entropy parameters, 
which are part of the free energy rules
derived from experiments (see \citet{walter} and
references therein).

{\bf Secondary structure.}
The number of possible secondary structures for a given sequence of
length $N$ grows exponentially with $N$.
To each structure ${\cal S}$, a Boltzmann weight $\zeta({\cal S})$ 
may be assigned with the
help of the free energy rules~\citep{walter} which
contain a large number of experimentally determined energy and
enthalpy parameters, e.g., those for the stacking of basepairs, formation of
internal, hairpin, bulge or multi-loops, and dangling ends.
Due to the large number of possible structures, the full partition function
$Z_N=\sum_{\cal S} \zeta({\cal S})$ is impossible to evaluate by enumeration, 
except for very small $N$. However, one can make use of recursion
relations that express the partition function for a subsequence with
the help of the partition functions for even shorter 
subsequences~\citep{mccaskill,zuker}, and proceed to compute the full
partition function exactly  in $O(N^3)$ time.  These recursion relations owe
their existence to the fact that the class of secondary structures
was defined to include only {\em nested} structures, e.g. two
basepairs $(i,j)$ and $(k,l)$ with $i<k<j<l$ are not admitted
(the occurrence of such pairings is called a pseudoknot and
contributes relatively little to the free energy of natural 
RNAs~\citep{tinoco}).
One implementation of this algorithm with very detailed free energy
rules is the `Vienna package' 
\citep[publically available at http://www.tbi.univie.ac.at/]{vienna}.
In the following, we describe the modifications that we made to this
package in order to obtain $\cQ_N(m)$
and the corresponding minimum free energy structures.

The Vienna package calculates the auxiliary partition function
$\Pi(i,j)$ for the substrand (i.e., a contiguous segment of 
the sequence) from base $i$ to base $j$,
under the condition that base $i$ and base $j$ are paired. These
quantities can be used to calculate the partition function
$Q(j;n)$ of the substrand from base $1$ to base $j$, under the
condition that the {\em exterior} part of the configurations is $0\le
n\le j$ bases long. The recursion formula for $Q$ 
is\footnote{Here, the constant $\Delta = 3$ accounts for the fact 
that each stem branching from the exterior single strand contributes 
an additional segment, whose length is approximately equal to the 
length of three single stranded bases.} 
\begin{displaymath}
Q(j\!+\!1;n)\!=\!Q(j;n\!-\!1)+
\!\!\!\!\!\!\!\!\sum_{i=n-\Delta+1}^j\!\!\!\!\!\!\!
Q(i\!-\!1;n\!-\!\Delta)\Pi(i,j+1),
\end{displaymath}
obtained by splitting the partition function $Q(j+1;n)$ up
according to all possible binding partners of base 
$j+1$. This formula,
together with the appropriate boundary conditions for $j=0$ and $n=0$,
can be solved recursively by calculating $Q(j;n)$ first for all $n$
at a given $j$ and then for increasing $j$.
In the end, we have $\cQ_N(m) = Q(N;m)$ 
for the $m$ exterior bases in $O(N^3)$ time.

To produce the minimum free energy structures at fixed $m$, we use
an equivalent recursive scheme, but replacing the summations by maxima
to obtain first the minimum free energy~\citep{zuker}.
Then, we determine the corresponding structure by going through the
scheme in reverse and reconstructing at each step which of the terms
was maximal.

{\bf Polymer model.} 
The simplest polymer model for the exterior single strand (the open 
circles in Fig.~\ref{figsketch}) is the Gaussian chain \citep{deGennes}. 
However, as shown below, the force-induced denaturation of RNA 
occurs at forces of order 10 pN, where the exterior single strand 
is strongly stretched and the Gaussian model breaks down. 
In this regime, an elastic freely jointed chain (EFJC) 
model\footnote{Self avoidance in the exterior single strand may be 
neglected, again because of its highly stretched state.}
yields a good fit to experimental FEC's \citep{smith,mezard}. 

The distance along the backbone between two adjacent nucleotides is
the segment length of the chain. We denote it by $l$ and assign an
elastic energy $V({\bf r})=\frac{\kappa}{2}(r-l)^2$ per segment, where
${\bf r}$ represents the end-to-end vector of the segment.  Instead of
attempting (the very cumbersome) exact computation of the end-to-end
vector distribution $W({\bf R};m)$ of the chain, we employ an
asymptotic expression that becomes exact in the limit of large $m$ and
is sufficiently accurate for our purposes even for small $m$.  It can
be derived along the line of a similar calculation for the case of the
regular (i.e. non-elastic) freely jointed chain given in
\citep{flory}.  The result is conveniently expressed in terms of the
quantity $q(h)=\int{\rm d}^3r\,e^{-{\bf h}\cdot{\bf r}-V({\bf
r})/k_{\rm B}T}/ \int{\rm d}^3r\,e^{-V({\bf r})/k_{\rm B}T}$, where
$k_{\rm B}$ denotes the Boltzmann constant and ${\bf h}$ is a vector
of length $h$ with fixed (but arbitrary) orientation in space.  The
asymptotic expression is then
\begin{equation}
  \label{Wm}
  W({\bf R};m)\approx C\frac{h}{2\pi R}\left[q(h)\right]^m
               e^{-hR}\;,
\end{equation}
where $C$ is a normalization constant and $h$ is determined from
$R=m\;\frac{\partial}{\partial h} \log q(h)$.
We incorporate the effect of a Kuhn length $b\!>\!l$ by rescaling
the end-to-end vector distribution through
$l\rightarrow b$ and $m\rightarrow ml/b$.

{\bf Observables.}
Apart from the force at fixed extension, which is calculated from
Eq.~(\ref{convolution}) by 
\begin{equation}
  \label{force}
  f(R)=- k_B T \, \frac{\partial}{\partial R}\log Z_N(R)
\end{equation}
(see, {\it e.g.} \citep{flory}), 
we also calculate the mean number of stems, $n_{\rm stem}$, along
the exterior chain (for the structure depicted in Fig.~\ref{figsketch}
this would be $n_{\rm stem}\!=\!2$).
This may be determined by introducing an extra free energy penalty,
$\varepsilon_{\rm stem}$, for each external stem into the calculation
of $\cQ_N(m)$ and then differentiating numerically with respect to
$\varepsilon_{\rm stem}$, i.e., 
\begin{displaymath}
  n_{\rm stem}(R)=-\left.k_B T \, \frac{\partial}
  {\partial\varepsilon_{\rm stem}}\log Z_N(R)
  \right|_{\varepsilon_{\rm stem}=0}\;.
\end{displaymath}

\begin{figure}[tb]
  \begin{center}
  \includegraphics[width=8cm]{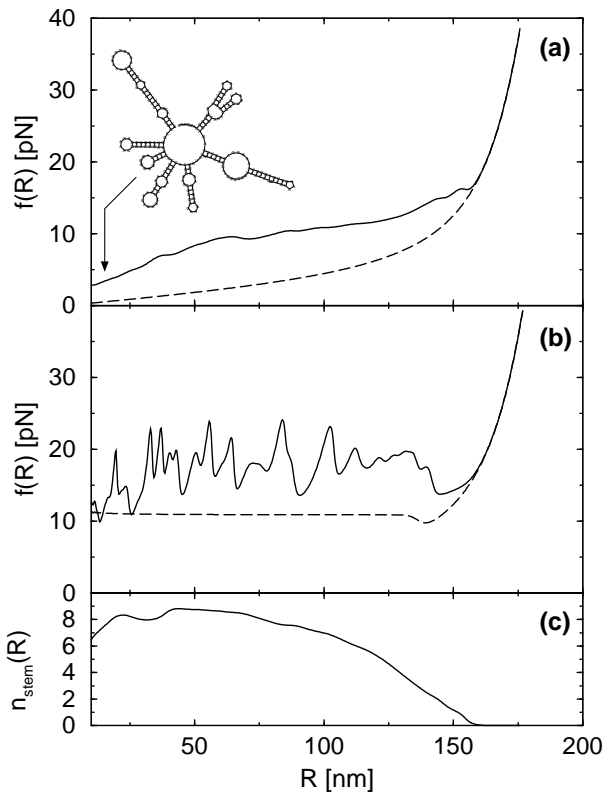}
  \end{center}
  \caption{ 
  (a) Force-extension curve (FEC) for a group I intron 
  (solid line, see text for details) and a homopolymeric RNA of the 
  same length, $N$=251 (dashed). 
  The depicted secondary structure is the minimum free energy 
  structure at $R=10nm$. 
  (b) FEC for a hairpin composed of randomly chosen basepairs (solid) 
  and a homogeneous hairpin of AU-basepairs (dashed). In 
  both cases the total sequence length is $N$=252. 
  (c) Mean number of exterior stems, $n_{\rm stem}(R)$, for the 
  group I intron.} 
  \label{figgroup1}
\end{figure}

{\bf Choice of Parameters.}
We work at room temperature, $T\!=\!20^\circ{\rm C}$, and use
the DNA polymer parameters obtained by \citet{mezard} by fitting 
to the experiment of \citet{bensimon} also for RNA, since we are 
not aware of the corresponding experimental data. (We do not expect 
a large difference in the single strand properties between DNA and
RNA, because of the high similarity between their chemical structures.) 
The values are
$l\!=\!0.7$nm, $b\!=\!1.9$nm, and $(\kappa/k_B T)^{-1/2}\!=\!0.1$nm.
We take the free energy parameters for RNA secondary structure 
as supplied with the Vienna package. 
The salt concentrations at which these free energy parameters were
measured are $[\rm{Na}^+]=1M$ and $[\rm{Mg}^{++}]=0M$.
\section{Results}
Fig.~\ref{figgroup1}a and b show the FEC's (solid lines) for two 
RNA sequences with practically the same total length and composition, 
both computed as described in the last section using the same set 
of parameters. 
Strikingly, the first curve is almost completely smooth with no 
significant features, while the second is extremely jagged with large 
`jumps' in the force. 
This dissemblance is entirely due to the difference between the 
secondary structures into which the two sequences fold. 
The sequence in Fig.~\ref{figgroup1}a originates from the group I 
intron of the methionine tRNA of {\it Scytonema hofmanii} with a 
sequence length of $N=251$ (GenBank\# U10481). 
Its dominant secondary structure (according to our 
algorithm\footnote{The known 
native secondary structure of this sequence contains two helical 
regions forming a pseudoknot. Since pseudoknots are excluded from 
our approach (as explained above), we removed it from the structure 
computationally by replacing 6 basepairs in the less stable of the 
two helical regions (positions no. 79--84 and 157--162) by artificial 
bases which are excluded from base pairing. With this modification, 
the minimum free energy structure at zero force (as determined by 
the Vienna package) is almost identical with the secondary structure 
known from comparative sequence analysis 
\citep[available at http://www.rna.icmb.utexas.edu/]{gutell} 
outside of the pseudoknot region. 
Beyond the distance at which the pseudoknot is pulled apart, our 
modification of the sequence should not effect the FEC significantly. 
This expectation is supported by our numerical observation that the 
FEC's for the unmodified sequence (ignoring the pseudoknot) and for 
our modified sequence become close to identical beyond a distance of 
$R\approx70~{\rm nm}$.}) at an extension of $R$=10nm is also depicted 
in Fig.~\ref{figgroup1}a. 
The sequence in Fig.~\ref{figgroup1}b was artificially generated by 
concatenating a randomly chosen sequence with its reverse complement, 
so that it folds into a single hairpin composed of random basepairs. 
Its FEC is very similar to the experimental force curve obtained upon 
unzipping double stranded DNA by \citet{heslot}; the sawtooth-like 
oscillations correspond to a `molecular stick-slip process' 
\citep{bockelmann}. 

Why does the group I intron not display an abundance of features in 
the FEC like the hairpin does? Its secondary structure consists of many 
structural elements (e.g. stem-loop structures), the opening of which 
one might expect to produce clear signatures in the FEC. 
Indeed, in their theoretical study of force-induced denaturation of 
DNA/RNA, \citet{siggia} concluded that the 
opening of individual basepairs in double stranded DNA cannot readily 
be observed, but the opening of stem-loop structures in RNA should be. 

One fairly obvious effect that could cause the smooth FEC is thermal 
superposition of alternative secondary structures. Since one may 
expect that typical RNA structures (such as the one depicted in 
Fig.~\ref{figgroup1}a) are less well-designed than a perfect hairpin, 
force-induced denaturation should make more alternative structures 
accessible in the former case than in the latter.  
In our analysis below, we find that this effect is indeed non-negligible, 
but the largest loss of features originates from another, 
more subtle mechanism, which we call the `compensation effect', and 
which persists even when no alternative secondary structures are allowed. 
The compensation effect depends on the fact that 
when several structural elements are pulled at {\em in parallel}, the 
optimization process that determines the minimum free energy structure 
with a given number $m$ of external open bases may {\em reclose} 
stretches of basepairs which had already been opened at a lower value 
$m'<m$. 

In our approach (see `Model and Methods' above), the information on 
the secondary structure energetics for a given sequence is entirely 
contained in the function $\cQ(m)$. With the help of the polymer model 
(contained in $W({\bf R};m)$) this information is translated into a 
FEC via Eq.~(\ref{convolution}). 
Our investigation therefore comprises two steps. 
First, we seek to understand what property of $\cQ(m)$ determines 
the size of the fluctuations in the FEC, and second, how this 
property depends on the secondary structure. 

\begin{figure}[tb]
  \begin{center}
  \includegraphics[width=5.2cm]{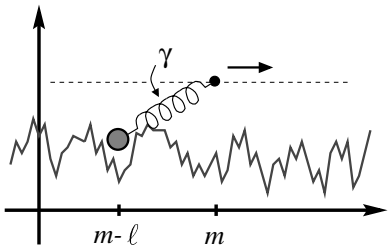}
  \end{center}
  \caption{ 
    The problem of RNA pulling (in the fixed-distance ensemble) 
    may be mapped onto the statistical mechanics problem of a particle 
    with a spring attached to it moving in a one-dimensional 
    disordered potential. The other end of the spring is externally 
    controlled and slowly advanced into one direction.}
  \label{figspring}
\end{figure} \noindent

The first question is addressed most readily for the special case of 
the random hairpin of Fig.~\ref{figgroup1}b. 
It is known that in the fixed-force ensemble, unzipping of a random 
hairpin may be mapped onto the problem of a particle in a tilted 
one-dimensional random potential \citep{deGennes75,lubensky}. 
The random potential is correlated and has the statistical 
properties of a one-dimensional random walk. 
In the fixed-distance ensemble, we may perform a very similar 
mapping\footnote{For 
these mappings, alternative structures of the hairpin sequence are 
neglected, which is a good approximation due to the perfect design 
of the hairpin. Also, the nearest-neighbour correlations in the 
random potential caused by the stacking energies are not taken into 
account, since they would not change the qualitative predictions of 
the model.} (see Fig.~\ref{figspring}). 
Here, the bias for the direction of movement of the particle is not 
caused by a tilt of the potential, but instead by a spring that is 
attached to the particle. The position of the other end of the 
spring is externally controlled, i.e. it is determined by $R$, the 
given end-to-end distance of the RNA molecule. 

\begin{figure}[tb]
  \begin{center}
  \includegraphics[width=6cm]{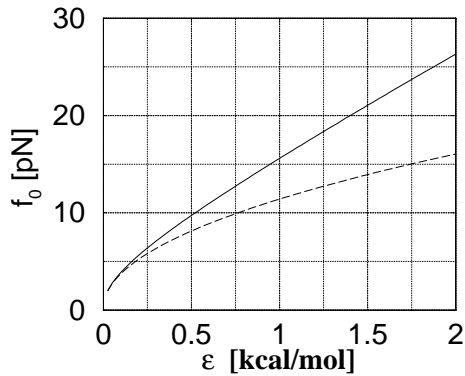}
  \end{center}
  \caption{ 
  Threshold force $f_0$ for unzipping of a homogeneous hairpin 
  as a function of the binding energy per base $\varepsilon$ (solid line). 
  The dashed line indicates the Gaussian approximation
  $f_0\!=\!(6k_BT\,\varepsilon/lb)^{1/2}$, which is 
  obtained by using the end-to-end distance distribution $W(R;m)$ of 
  a Gaussian chain. Note that the Gaussian approximation breaks down 
  already at low forces, and the more detailed treatment according to 
  Eq.~(\ref{Wm}) is necessary. 
  The stacking energy for  AU-pairs at $T\!=\!20^\circ{\rm C}$ is 
  $2\,\varepsilon\!\approx\!1.21~{\rm kcal/mol}$ corresponding to a 
  threshold force $f_0\approx 11~{\rm pN}$, which agrees with the value 
  observed in Fig.~\ref{figgroup1}b (dashed line).}
  \label{figfc}
\end{figure}

In the following, we review the relation between the parameters of the 
particle-in-a-random-potential problem, i.e. the spring constant 
$\gamma$ and the variance of the random potential, and the parameters 
of the unzipping problem. 
This will also serve us to introduce our notation for the subsequent 
discussion. 
We may write the free energy $G(m)=-k_BT\log\cQ(m)$ of the random 
hairpin as $G(m)=-\sum_{i=1}^{N-m}\eta(i)$, where the $\eta(i)$ are 
random with mean $\langle\eta\rangle=\varepsilon$ and variance 
$\langle\eta(i)\eta(j)\rangle-\langle\eta\rangle^2=
 \delta_{ij}(\Delta\varepsilon)^2$. 
Here, $\varepsilon$ represents the mean binding energy per base, which 
depends on the GC-content of the hairpin, the temperature, and the salt 
concentrations, and $\Delta\varepsilon$ measures the fluctuations of 
$\varepsilon$, both along a given hairpin and between different 
realizations of the random sequence. 
The difference between two free energies that are $\ell$ units apart, 
$\Delta G(\ell)=G(m)-G(m\!-\!\ell)$, then has the variance 
\begin{equation}
  \label{varDG}
  {\rm var}(\Delta G(\ell))=\ell\,(\Delta\varepsilon)^2\;.
\end{equation}
In the particle picture (see Fig.~\ref{figspring}), $m-\ell$ corresponds 
to the position of the particle, and $m$ to the position of the other 
end of the spring. For fixed $m$, the particle therefore sees the 
effective potential
\begin{equation}
  \label{Veffective}
  \Delta G(\ell)+\frac{\gamma}{2}\,\ell^2\;,
\end{equation}
i.e. Eq.~(\ref{varDG}) determines the variance of the random potential. 
The spring constant $\gamma$ is determined by $\varepsilon$ as follows. 
If $\Delta\varepsilon$ were zero, the unzipping force would take a 
constant value $f_0$ (cf. the dashed line in Fig.~\ref{figgroup1}b, which 
shows the FEC of a homogeneous AU-hairpin).
The dependence of $f_0$ on $\varepsilon$ can be calculated analytically 
by evaluating the sum in Eq.~(\ref{convolution}) by the saddle point 
method (see also (D.K.~Lubensky and D.R.~Nelson, in preparation)). 
The result is shown in Fig.~\ref{figfc} (solid line). 
Now $\gamma=l^2\Gamma$, where $\Gamma$ is the local spring 
constant of a nonbinding RNA of $m$ bases at force $f_0(\varepsilon)$. 
Since the spring constant of a homopolymer scales with the inverse of the 
number of segments, we write $\Gamma\!=\!\Gamma_0/m$, where $\Gamma_0$ 
depends only on $f_0$, but not on $m$. 
Graphically, $\Gamma_0(f_0)$ is the slope at $f\!=\!f_0$ of the 
dashed line in Fig.~\ref{figgroup1}a (FEC of a homopolymeric 
RNA), multiplied by 251 (the number of bases in that example). 
In this way $\Gamma_0(f_0)$ may also be determined from an experimental 
FEC. 

 \noindent

When the fluctuations in the random potential are not too weak, the 
particle follows the other end of the spring in {\em discrete jumps}. 
The typical size of a jump, $\Delta\ell_{\rm jump}$ is given by the 
value of $\ell$ for which the two terms in Eq.~(\ref{Veffective}) are 
of equal size, 
$\Delta \ell_{\rm jump}\simeq(2m\,\Delta\varepsilon/l^2\Gamma_0)^{2/3}$. 
A typical jump then leads to a drop in the force by 
$\delta{f}\simeq\Gamma l\,\Delta\ell_{\rm jump}$, i.e.
\begin{equation}
  \label{df}
  \delta f\simeq\left(4\,\Gamma_0\,\Delta\varepsilon^2/R\right)^{1/3}\;.
\end{equation}
This is valid as long as the thermal broadening of the particle 
position, $\Delta\ell_{\rm T}\simeq(2m/l^2\Gamma_0\beta)^{1/2}$, is less 
than the typical jump size $\Delta\ell_{\rm jump}$. 
In the opposite case, the particle is {\em sliding} more or less 
smoothly, and $\delta f\propto\Delta\varepsilon$.  

Eq.~(\ref{df}) furnishes an estimate for the size of the fluctuations 
in the FEC {\em for the case of a random hairpin}. 
However, since we used an arbitrary function $G(m)$ as input, the above 
argument may be made in general {\em for any structure}, as long as 
Eq.(\ref{varDG}) holds sufficiently well. 
Alternatively, if for a particular structure the dependence of 
${\rm var}(\Delta G(\ell))$ on $\ell$ is determined numerically, this could 
be used to replace Eq.~(\ref{varDG}) and Eq.~(\ref{df}) would have to be 
modified accordingly. 

\begin{figure}[tb]
  \begin{center}
  \includegraphics[width=8cm]{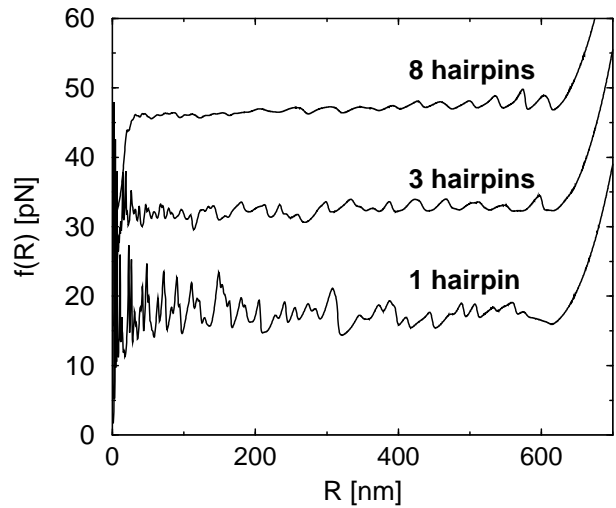}
  \end{center}
  \caption{ 
  Force-extension curves for 1, 3, and 8 hairpins with random 
  basepair composition in a row (sequence length $N\!=\!1000$; 
  the middle and upper curves are vertically shifted by 15 and 
  30 pN respectively). 
  Clearly, the fluctuations in the force curve decrease with increasing 
  number of hairpins, except for the last third of the extension 
  interval, where some of the hairpins of the 8 hairpin curve have 
  already completely disappeared. In our analysis described in the main 
  text only the first two thirds of all FEC's were used. 
  The decrease of the force fluctuations with increasing extension is due 
  to the entropic elasticity of the exterior single strand as described 
  by the $R$-dependence in Eq.~(\protect\ref{df}).} 
  \label{fig1to8hps}
\end{figure}

\begin{figure}[tb]
  \begin{center}
  \includegraphics[width=6.5cm]{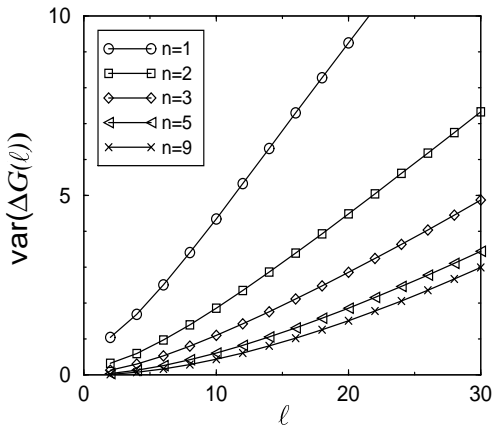}
  \end{center}
  \caption{ 
  The variance of $\Delta G(\ell)$ for different numbers of 
  hairpins.} 
  \label{fig_varDG}  
\end{figure}

We now address the question of how the fluctuations in $G(m)$ depend on 
the secondary structure. 
An essential difference between unzipping of a hairpin and 
force-induced denaturation of a typical RNA structure is that in the 
latter case, several stems are being pulled on simultaneously\footnote{In 
principle, a situation where several stems are pulled on in parallel can 
also arise in the process of unzipping a single long hairpin, due to 
accidental palindromic regions in the single strand which has already been 
pulled out. However, these non-native interactions have to overcome the 
energetic advantage of the native single-hairpin interactions, in order for 
the effect to become relevant. Hence the palindrome needs to be extremely 
GC-rich. For a single hairpin consisting of random basepairs, we estimated 
that a non-negligible palindrome would typically occur only in sequences of 
at least several thousand bases in length, which is beyond the length of 
the sequences studied here.}
for most of the extension interval (see Fig.~\ref{figgroup1}c, which shows 
the number of stems as a function of the extension for the group I intron 
studied above). 
To analyze the effect of multiple stems, we constructed artificial 
sequences that form a given number $n$ of random hairpins in a row 
(i.e. the sequences are a concatenation of $n$ random hairpin sequences, 
each of which is constructed as explained above). 
For each $n$ in the range $1\le n<10$, we computed $G(m)$ and the FEC's 
for 1000 different sequence realizations, all with an 
approximate total length of $N\!=\!1000$.  
As an example, Fig.~\ref{fig1to8hps} shows the FEC's for three 
sequences, which fold into $n\!=\!1$, 3, and 8 hairpins, respectively. 
Clearly, the fluctuations in the force curve decrease with increasing $n$. 
We obtained ${\rm var}(\Delta G(\ell))$ as an average over the 1000 
realizations and a small interval of $m$. 
Some of the resulting curves are shown in Fig.~\ref{fig_varDG}. 
Although the dependence of ${\rm var}(\Delta G(\ell))$ on $\ell$ is not 
completely linear, the deviation from linearity over the small range of 
$\ell$-values relevant here (typically, $0\le\ell\le 12$) is not very large. 
For the sake of simplicity, we chose to interpret the data with the 
theory for a linear ${\rm var}(\Delta G(\ell))$ developed above. 
To this end, we define an effective $\Delta\varepsilon$ for each $n$ 
from the slope of ${\rm var}(\Delta G(\ell))$ at $\ell\!=\!4$. 

\begin{figure}[tb]
  \begin{center}
  \includegraphics[width=5cm]{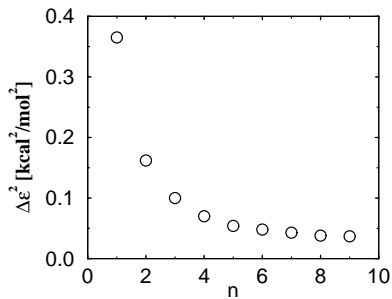}
  \end{center}
  \caption{ 
  Dependence of $\Delta\varepsilon^2$ on the number of 
  hairpins (circles).}
  \label{depssq_vs_n} 
\end{figure}

Fig.~\ref{depssq_vs_n} shows that $\Delta\varepsilon^2$ decreases 
monotonically with the number of stems that are being pulled on 
simultaneously. 
This decrease is almost entirely due to the compensation effect, 
which we may intuitively understand as follows. 
When a single hairpin is being unzipped, the stick-slip process 
described in \citep{heslot} is topologically inevitable, since 
the basepairs have to be opened in the order they occur. 
A strongly bound region that is followed by a weakly bound one,  
then always leads to a rise and subsequent drop of the FEC. 
However, with several hairpins, only the total number of exterior 
open bases is externally constrained, while the individual hairpins 
may freely open and reclose basepairs (for equilibrium FEC's there 
is no kinetic constraint). 
Therefore, if in a particular hairpin a strongly bound region is 
followed by a weakly bound one, both regions can open together and 
another hairpin can reclose a few basepairs to compensate for the 
released single-strand. 
Obviously, with a growing number of hairpins this mechanism will 
be increasingly effective. 
Clearly, in the fixed-force ensemble, the compensation effect is 
equivalent to an average over the FEC's of the individual hairpins. 
Moreover, with a large number of hairpins, the fixed-force and the 
fixed-distance ensembles become equivalent 
(D.K.~Lubensky and D.R.~Nelson, in preparation). 

\begin{figure}[tb]
  \begin{center}
  \includegraphics[width=7cm]{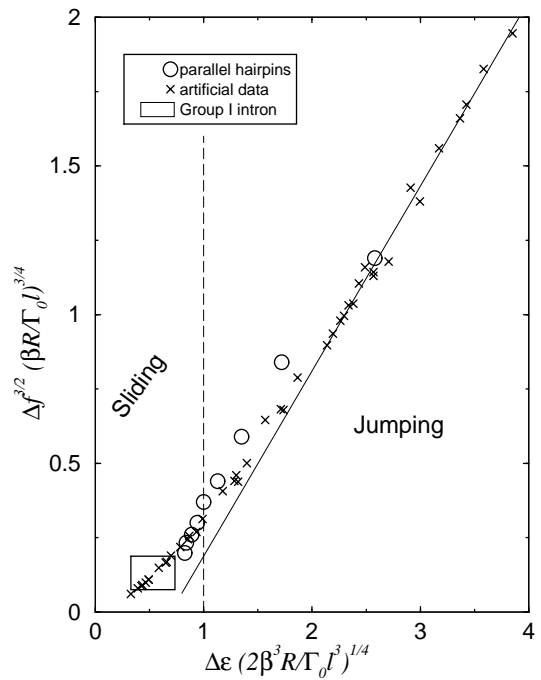}
  \end{center}
  \caption{ 
  Scaling plot of the force fluctuations against the free 
  energy fluctuations. The dashed vertical line marks the crossover 
  region between the jumping regime and the sliding regime. The solid 
  line is a linear fit to the data with abscissae larger than two. 
  It confirms the scaling behavior expected for the jumping regime. 
  See text for details.} 
  \label{df_scaling}
\end{figure}

To analyze the force fluctuations quantitatively, we calculated the 
FEC's for all of the 1000 sequence realizations of the $n$ parallel 
hairpins, and defined $\Delta{f}(R)$ as the standard deviation of the 
force at extension $R$ (the so-defined $\Delta{f}$ is smaller than 
the typical size of a force jump, $\delta{f}$, but should have the 
same scaling behavior). 
Fig.~\ref{df_scaling} shows a plot of the force fluctuations against 
the free energy fluctuations, where the horizontal axis,
$\Delta\varepsilon(2\beta^3 R/l^3\Gamma_0)^{1/4}=
(\Delta\ell_{\rm jump}/\Delta\ell_{\rm T})^{3/2}$, 
is scaled such that it separates the jumping regime from the sliding 
regime at a crossover value of one. 
The vertical axis is scaled such that the data should collapse onto a 
straight line in the jumping regime according to Eq.~(\ref{df}). 
In order to guide the eye, Fig.~\ref{df_scaling} also displays 
artificial data (crosses) for which $G(m)$ was generated by drawing 
random numbers $\eta(i)$ and taking $G(m)=-\sum_{i=1}^{N-m}\eta(i)$ 
(the different points are for different values for the mean and 
variance of $\eta(i)$).
The circles mark the data points for the parallel hairpins and the 
rectangular symbol in the lower left indicates in what region the 
group I intron is situated\footnote{The rectangular area marks the 
range of points that we obtained by determining $\delta{f}$,  
$\Delta\varepsilon$, and $\Gamma_0$ by averaging over different 
extension intervals, all within the range 50--110 nm, which is a 
region where the mean force is relatively constant 
(this is required in order to separate fluctuations in the force from 
a gradual change in the mean value).}. 

For the artificial data (crosses) the above scaling arguments should 
apply rigorously. 
Indeed, the artificial data falls onto a straight line in the jumping 
regime (the solid line represents a linear fit to the points with 
abscissae larger than two) and in the sliding regime $\Delta f$ is 
proportional to $\Delta\varepsilon$ (not shown). 
For the real data, Fig.~\ref{df_scaling} shows that passing from a 
single hairpin through structures with several parallel perfect 
hairpins to a typical natural RNA may be viewed as passing 
from the jumping regime to the sliding regime for a particle in a 
(correlated) random potential. At the same time, the FEC's change 
from jagged to smooth. 

\begin{figure}[tb]
  \begin{center}
  \includegraphics[width=7cm]{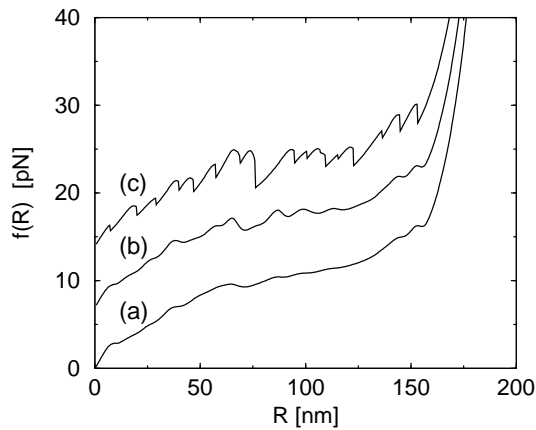}
  \end{center}
  \caption{ 
  Force-extension curves for the group I intron under different 
  conditions. Curve (a) is a copy of the full thermodynamic curve of 
  Fig.~\ref{figgroup1}a. 
  Curve (b) (vertically shifted by 7pN) was calculated by taking only 
  the minimum free energy structures along the unfolding pathway into 
  account, i.e. the thermal smoothening due to suboptimal structures 
  is suppressed. 
  For curve (c) (vertically shifted by 14pN), the rebinding of basepairs 
  that had already opened at smaller extension has also been suppressed, 
  in order to simulate non-equilibrium pulling. 
  } 
  \label{fast_intron}
\end{figure}

As mentioned above, thermal superposition of alternative secondary 
structures also contributes to the smoothening of the FEC's: 
since the structural elements in each suboptimal structure  
open at different values of $m$, the thermal average over all these 
structures smoothes $G(m)$.  
In order to assess the importance of this effect, we suppressed it by 
taking only the minimum free energy secondary structures into account 
instead of calculating the full partition function $\cQ(m)$.   
For the group I intron, the FEC without the contribution of suboptimal 
structures is shown in Fig.~\ref{fast_intron}b. 
Compared to the full thermodynamic curve (shown in Fig.~\ref{fast_intron}a), 
some structure is gained, but not nearly as much as in 
the FEC for the random hairpin of the same length, Fig.~\ref{figgroup1}b. 
This indicates that the compensation effect is the dominant source for 
the smoothing of the FEC. 
\section{Discussion}
In the last section, we found that the equilibrium FEC's for typical 
RNA molecules (like the group I intron that served us as an example) are 
quite smooth and do not reveal any features that can be associated with 
the opening of structural elements. 
The compensation effect is the primary cause for this result, and we 
expect it to be responsible, in part, also for the experimental 
observation of extremely smooth FEC's for single-stranded DNA by 
\citet{bensimon}. 
Nevertheless, the measurement of equilibrium FEC's for RNA or 
single-stranded DNA might still be useful, e.g. for an experimental 
determination of the RNA/DNA free energy parameters. 
Usually, these are extracted from melting curves of oligomers \citep{freier}, 
which requires variation of the temperature away from the temperature of 
interest up to the melting point of the oligomers, where the free energy and 
its temperature derivative are determined. 
The free energy parameters at the temperature of interest are then obtained 
by extrapolation, which introduces an error inherent to the method.  
For pulling experiments, the temperature can be kept constant at the value 
of interest, which is an obvious advantage. 
Here, the limiting factor is only the precision of the force measurement. 
The quantitative relationship between stacking energy and threshold force 
expressed by Fig.~\ref{figfc} furnishes the necessary link between force 
and energy. 
Measuring FEC's for periodic hairpins composed of different building 
blocks, would lead to curves like the dashed line in Fig.~\ref{figgroup1}b 
with different values for the threshold force. 
From these values, the stacking energies could then be determined, 
which might lead to more accurate parameters at the desired temperature 
and salt concentrations.  

There are (at least) two options to obtain FEC's with more features, which 
in turn might allow one to obtain information on RNA secondary structure 
from pulling experiments. 
One could either record {\em non-equilibrium} FEC's or analyze the 
{\em fluctuations} around the equilibrium curve. 
For our theoretical investigation, the latter option is not available as 
long as we work in the fixed-distance ensemble, since the force 
fluctuations around the thermodynamic average diverge in that ensemble. 
We will pursue this option in a separate publication by working in 
a mixed ensemble (Gerland, U., R. Bundschuh, and T. Hwa, in preparation).
Here, we briefly consider non-equilibrium FEC's, 
where the rate of external increase in the force/extension is 
higher than (some of) the rates associated with internal 
rearrangements in the secondary structure. 
In the case of long {\em proteins}, either naturally occurring 
as an array of globular domains \citep{rief97} or synthesized 
protein arrays \citep{yang}, mechanical stretching experiments 
resolved the unfolding of up to 20 individual domains. 
These experiments were performed under non-equilibrium 
conditions \citep{rief98} with typical pulling speeds of 
$1 \mu$m/s.

\begin{figure}[tb]
  \begin{center}
  \includegraphics[width=8.5cm]{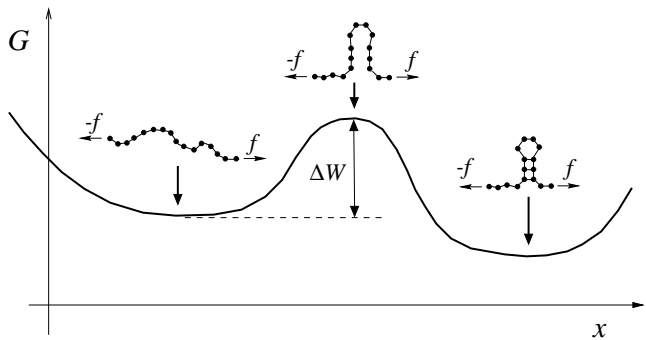}
  \end{center}
  \caption{ 
  Sketch of the assumed pathway for the formation of a stem-loop 
  structure in the presence of a stretching force $f$. 
  A generalized reaction coordinate $x$ is plotted along the horizontal 
  axis and the free energy $G$ along the vertical axis.  
  The work that has to be exerted against the force in order to pull in 
  the single strand needed for the formation of the stem-loop structure 
  is denoted by $\Delta W$. In principle, the entropy difference between 
  the random coil state on the left and the transition state also 
  contributes to the barrier height, however, we assume that at typical 
  stretching forces it is negligible compared to $\Delta W$.} 
  \label{loop_formation}
\end{figure}

In order to estimate whether non-equilibrium conditions are 
attainable for RNA with reasonable pulling speeds, we need a 
rough idea of the timescales involved in secondary structure 
rearrangements of RNA. 
For this, we again assume that RNA and single-stranded DNA behave 
similarly, so that we may draw on an experiment by 
Bonnet, Krichevsky, and Libchaber \citep{bonnet} 
measuring the opening and closing rates of DNA stem-loops using 
fluorescence correlation spectroscopy. 
From their results, we extract $\rm 10\,\mu s$ as an estimate for 
the closing time (at $T\!=\!20^\circ{\rm C}$) of a stem-loop 
structure with three basepairs and a loop of four nucleotides, which 
may be considered as a minimal secondary structure element. 
We expect that the formation of the stem-loop takes place in a single 
step whose reaction pathway goes through a transition state 
where the basepairs of the stem have not yet formed, but the 
corresponding bases are already closely together 
(see Fig.~\ref{loop_formation}). 
In the presence of an external force, the closing time must then be 
multiplied with an Arrhenius factor $e^{\Delta W/k_BT}$, where 
$\Delta W$ is the work that has to be exerted against the force to 
pull in the amount of single strand needed for the formation of the 
stem-loop \citep{rief98}. 
With a typical force of $\rm 6\,pN$ we obtain 
$\Delta W\!\approx\!4\,{\rm kcal/mol}$, which results in a 
closing time on the order of ${\rm 10\,ms}$. 
This timescale has to be compared to the time it takes to stretch out 
the stem-loop. 
At a pulling speed on the order of $1 \mu$m/s, the two timescales are 
comparable and hence, both the formation of new secondary structure 
elements and the restoration of already opened ones are likely to be 
suppressed\footnote{This estimate {\em does not apply} for the 
{\em rezipping} of partially opened, perfectly complementary long 
hairpins, which is faster than closing of a stem-loop. However in real 
RNA structures, long stems are usually interrupted by internal or bulge 
loops, which we expect to reclose on similar timescales as the 
stem-loops.}. 
Although it is beyond the scope of this paper, we want to note that in 
the presence of pseudoknots and/or tertiary interactions, 
the formation or re-formation of structural elements is expected to be 
slowed down even further, due to long search times for the interaction 
partners. 

To obtain an impression of how many features a non-equilibrium FEC 
might show for the group I intron, we change our equilibrium algorithm, 
such that the rebinding of bases is disabled once they have been 
unbound, and include only the contribution of the minimum free energy 
structures instead of all possible secondary structures. 
This is clearly a very crude approximation. 
In a proper treatment, only those kinetic processes whose energy 
barrier is higher than a certain threshold as determined by the 
pulling speed should be suppressed. 
Also, we did not account for the fact that the opening of basepairs 
occurs at higher forces in non-equilibrium as a consequence of 
Kramers theory \citep{evans}. 
Nevertheless, the FEC shown in Fig.~\ref{fast_intron}c gives an idea 
of the large number of structural transitions that take place during 
force-induced denaturation (for comparison, the equilibrium FEC is 
shown again in Fig.~\ref{fast_intron}a). 
We therefore believe that non-equilibrium stretching experiments of 
RNA could lead to interesting and useful results. 

We made most of the software tools developed for the present work 
available to the public by creating a 
`RNA pulling server' at 
{\tt http://bioinfo.ucsd.edu/RNA}.
%
\vspace*{3mm}\newline\noindent
We thank D.~Bensimon, C.~Bustamante, and J.D.~Moroz for stimulating 
discussions. 
U.G. is supported by the Hochschulsonderprogramm III of the 
DAAD. R.B. and T.H. acknowledge support by the NSF through Grant 
No. DMR-9971456, DBI-9970199, and the Beckmann foundation. 
\end{document}